\documentstyle[prl,aps,multicol,epsf]{revtex}
\begin{document}

\title{Disorder and Funneling Effects on Exciton Migration in Tree-Like
Dendrimers}
\author{Subhadip Raychaudhuri$^{(1)}$, Yonathan Shapir$^{(1)}$,
 and Shaul Mukamel$^{(2)}$}
\address{$^{(1)}$Department of Physics and Astronomy, University of
Rochester, Rochester, NY 14627\\ $^{(2)}$Department of Chemistry,
University of Rochester, Rochester, NY 14627}
\date{\today}
\maketitle

\begin{abstract}
The center-bound excitonic diffusion on dendrimers subjected to several
types of non-homogeneous funneling potentials, is considered. We first
study the mean-first passage time (MFPT) for diffusion in a linear
potential with different types of correlated and uncorrelated random
perturbations. Increasing the funneling force, there is a transition
from a phase in which the MFPT grows exponentially with the number of
generations $g$, to one in which it does so linearly. Overall the
disorder slows down the diffusion, but the effect is much more
pronounced in the exponential compared to the linear phase. When the
disorder gives rise to uncorrelated random forces there is, in
addition, a transition as the temperature $T$ is lowered. This is a
transition from a high-$T$ regime in which all paths contribute to the
MFPT to a low-$T$ regime in which only a few of them do. We further
explore the funneling within a realistic non-linear potential for
extended dendrimers in which the dependence of the lowest excitonic
energy level on the segment length was derived using the Time-Dependent
Hatree-Fock approximation. Under this potential the MFPT grows
initially linearly with $g$ but crosses-over, beyond a
molecular-specific and $T$-dependent optimal size, to an exponential
increase. Finally we consider geometrical disorder in the form of a
small concentration of long connections as in the {\it small world}
model. Beyond a critical concentration of connections the MFPT
decreases significantly and it changes to a power-law or to a
logarithmic scaling with $g$, depending on the strength of the
funneling force.
\end{abstract}

%Submit to Physical Review E

\section{INTRODUCTION}

Extended dendrimers (Fig~1.) are nanoscale Cayley tree like
supermolecules which exhibit an energy gradient from the periphery to
the center of the molecule. Their unique hierarchical self-similar
structure can be described by three ingredients: the basic building
block (e.g. phenyl acetylene linear segment), the branching of the end
points and the number of generations. The number of basic elements
grows exponentially with the number of generations g and the number of
elements at the periphery of the molecule is comparable to the number
of bulk elements.

Recent theoretical and experimental studies \cite{kopel} \cite{tomal}
\cite{blazani} \cite{frechet} have shown that the electronic
excitations of these dendrimers are spatially localized within each
segment. For these dendrimers, lengthening of the linear tree branches
towards the core leads to a hierarchy of localization lengths. Hence
the exciton energy decreases with generation from the periphery to the
core. This {\em energy funnel} combined with the large number of
absorbing elements at the periphery make these dendrimers potentially
efficient single molecule antenna systems \cite{devadoss}. Excitons
created upon optical excitation of the shortest linear segments at the
periphery, diffuse through the intermediate regions, finally reaching
the core where an energy trap is located. Excitation transfer proceeds
via Coulomb interaction and may be described by the Frenkel exciton
model. Theoretical studies \cite{klafter1} focused on calculating the
time it takes for an exciton generated at the surface to reach the
core. This trapping time is a direct measure of the efficiency of the
antenna. Its dependence on the number of generations $g$ and the
funneling driving force were calculated.

These studies which were able to capture some important features of
excitonic diffusion on dendrimers, assumed that all the tree branches
(namely the linear polymeric chains) of a given generation are
identical. This implies that the exciton energies are fixed. In reality
interaction with the solvent and intramolecular vibrations induce slow
(quenched) fluctuations in the energy $\epsilon$. (Nonlinear
\cite{rella} and single molecule \cite{hplu} spectroscopy typically
show nanosecond to millisecond bath motions whereas the exciton
trapping times are typically in the picosecond range \cite{devadoss}
\cite{pope}).

On a more fundamental level, diffusion in disordered media is an active
field of Statistical Physics\cite{24}. Different forms of correlations
among energy levels were considered and nontrivial behaviors, such as
the difference between the typical and the average diffusion, anomalous
scaling, and others were found in some cases. Independently,
theoretical investigations of dynamics on Cayley trees (or Bethe
lattices) lead to many interesting results. Exact solutions exist for
some problems and bear important consequences. Directed polymer on
Cayley trees is one of the interesting examples where a spin-glass like
transition was predicted \cite{derrida}. As shown below the problem of
diffusion on a disordered Cayley tree exhibits such a transition as
well, and thus has a considerable theoretical significance.

In a recent letter \cite{subha}, we reported preliminary studies of the
effects of nonlinear dependence of the excitonic energy and realistic
quenched disorder on excitonic diffusion on Cayley tree like
dendrimers. We further carried out analytic investigations of diffusion
on a disordered Cayley tree with a linear potential. Our findings can
be summarized as follows:

(i) Due to the non-linear variation of the funneling potential with the
generation, an optimal the generation number $n^*(g)$ was found beyond
which the nonlinear potential drastically diminishes the
light-harvesting efficiency. Moreover, increasing of the trapping time
by disorder (by slowing down the exciton diffusion) will be more
pronounced if the dendrimer crosses this optimal size. Hence, for
dendrimers larger than the optimal size, even though the total photon
absorbance will increase due to the increase in the number of
peripheral light absorbing sites, the slower excitonic migration
towards the active center will make the light-harvesting antenna less
effective.

(ii) We investigated the diffusion on a Cayley tree in the presence of
a linear potential and considered various types of disorder, depending
on the correlations among the energy fluctuations. For a specific type
of disorder (See model (iv) below), we found a dynamic phase transition
to a highly disordered phase where only a few paths dominate the
exciton migration. This resembles the equilibrium replica symmetry
breaking transition found in other random systems (e.g. directed
polymers in random media).

The purpose of this paper is to elaborate on, and extend the results
reported in our previous study. We further report the numerical study
of diffusion on a Cayley tree with a specific kind of
geometric disorder obtained by adding a few random connections between
various sites to the usual tree  structure. This {\em small world}
model was first introduced by Watts and Strogatz \cite{watts} for
Euclidean lattices.

The organization of the paper is as follows: The MFPT  for random walk
is reviewed in Sec. II. In Sec. III we discuss diffusion on a Random
Cayley tree with a linear potential. The effect of nonlinear potential
on the exciton migration is reported in Sec. IV.  The small world model
is studied in Sec. V. Finally, we conclude in Sec. VI with a summary of
our results.

\section{RANDOM WALKS AND THE MEAN FIRST PASSAGE TIME (MFPT)}

The Mean First Passage Time (MFPT) is a faithful measure of the
trapping efficiency defined as the average time it takes for an exciton
to diffuse from the periphery to the core, where it gets trapped
\cite{27}. Another quantity of interest related to the MFPT is the Mean
Residence Time (MRT), which is the average time spent on a site of the
tree (branch of the dendrimer)[13,14,23,24]. Our theoretical effort
focused on calculating the MFPT and the MRT for a continuous time
random walk with exponential distribution of waiting times which gives
a master equation for the probability $P_{n}(t)$ of the exciton to be
on site $n$ at time $t$. If the energy depends only on n, the problem
becomes effectively one-dimensional for $P_{n}(t)$. The random walker
on the $n$-th site of the one-dimensional chain jumps to nearest sites
with rates $R_{i}$ (toward the reflecting point) and $T_{i}$ (toward
the trap). The random walk is described by the master equation
\cite{montroll}

\begin{equation} \label{eq:}
    \frac{d{\bf P}(t)}{dt} =- A {\bf P}(t),
\end{equation}
where ${\bf P}(t)$ denotes the probability vector and the survival
probability $S(t) = \sum _{1}^{g} P_{n}(t)$. $A$ is the tridiagonal
transition matrix. Written explicitly in terms of the rates $R$ and
$T$:

\begin{eqnarray}
\dot{P}_{0}(t) & = & T_{1}P_{1}(t),  \nonumber   \\ \dot{P}_{1}(t) & =
& T_{2}P_{2}(t) - (T_{1}+R_{1})P_{1}(t), \nonumber \\ \dot{P}_{n}(t) &
= & T_{n+1}P_{n+1}(t) + R_{n-1}P_{n-1}(t) - (T_{n}+R_{n})P_{n}(t)
\qquad        (1<n<g),    \\ \dot{P}_{g}(t) & = & R_{g-1}P_{g-1}(T) -
T_{g}P_{g}(t). \nonumber
\end{eqnarray}

The trap $(P_0)$ is not included in Eq. (1), while reflecting Boundary
Condition are imposed at $n=g$.

The formal solution of Eq. (1) is

\begin{equation} \label{eq:}
       {\bf P}(t) = \exp (-At){\bf P}(0),
\end{equation}
where ${\bf P}(0)$ is the initial condition. ${\bf P}(0) = [0,0,0
....1,0,0,0...]^{T}$ for an excitation which starts on the $n$-th
generation. The MFPT is obtained by integrating the survival
probabilities $P_{n}(t)$ over time

\begin{equation} \label{eq:}
 \tau(g) = \int_{0}^{\infty}  \sum_{n=1}^{g} P_{n}(t) dt
         = \sum_{n=1}^{g}  \int_{0}^{\infty} P_{n}(t)
         \equiv \sum_{n=1}^{g} t_{n},
\end{equation}
where $t_{n}$ is the MRT for generation $n$. Using the general solution
Eq. (3)  \cite{montroll} \cite{kampen}

\begin{equation} \label{eq:}
 \int_{0}^{\infty} {\bf P}(t) dt = \int_{0}^{\infty} \exp (-At) {\bf P}(0) dt
                           =  A^{-1} {\bf P}(0).
\end{equation}

Hence the MRT at site $n$ is given by

\begin{equation} \label{eq:}
  t_{n} = \int_{0}^{\infty} P_{n}(t) = \sum_{j} A^{-1}_{nj} P_{j}(0).
\end{equation}

For an excitation which starts at the highest (periphery) generation
$g$, $ P_{j}(0) = \delta _{jg}$ and the MRT is given by
\cite{klafter1}

\begin{eqnarray}
    t_{1} & = & \frac{1}{T_{1}}   \hspace{5.5cm}   n=1  \\
    t_{n} & = &  \frac{1}{T_{n}} + \sum_{i=0}^{n-2}
(\frac{1}{T_{i}}) \prod_{j=i+1}^{n-1} \xi_{j} \hspace{2cm} 1< n \le g
\end{eqnarray}

where $\xi_{n} \equiv R_{n}/T_{n+1}$ is the detailed balance ratio. The
MFPT is given by

\begin{eqnarray}
 \tau(g) & = & \sum_{n=1}^{g-1} t_{n}     \\
         & = & \sum_{m=1}^{g} \frac{1}{T_{m}} + \sum_{m=2}^{g-2}
    \sum_{i=0}^{m-2} (\frac{1}{T_{i}}) \prod_{j=i+1}^{m-1} \xi_{j}.
\end{eqnarray}

An explicit form of this equation is given in the appendix. Hereafter
we assume $T_{n}=1$ and treat $\xi_{n}$ as independent variables with
$\xi_{n} = c \exp \{-\beta(\epsilon(n+1)-\epsilon(n))\}$ where $c +1$
is the coordination number of the tree ($c=2$ in our case),
$\epsilon(n)$ denotes the energy of the nth segment and
$\beta=(KT)^{-1}$ is the Boltzmann factor.

\section{RANDOM WALK AND MFPT ON A DISORDERED  CAYLEY TREE WITH A
LINEAR Funnel}

The linear funneling potential assumes the form
\begin{equation}
\overline{\epsilon}(n)= nF
\end{equation}
$F$ being the potential difference. For this model $\xi_{n} = \xi_{0}$
= $c \exp{(-\beta F)}$. Eq. (10) can be easily summed (see the
appendix for details) to yield for the MFPT

\begin{equation} \label{eq:}
\tau (g) =  \left \{ \begin{array}{ll}
  \xi \frac{\xi^{g} -1}{(\xi -1)^{2}} -
      \frac{g}{\xi -1}  \quad &  \textrm{for $\xi \neq 1$}  \\
  \frac{g(g+1)}{2}     \quad &  \textrm{for  $\xi = 1$}.
           \end{array} \right.
\end{equation}

For large g ($\gg 1$) Eq. (12) shows three distinct behaviors

(a) $\xi > 1$: $\tau(g) \sim \exp (g \ln \xi)$  (exponential regime).

(b) $\xi < 1$: $\tau(g) \sim g$ \hspace{0.5cm}(linear regime).

(c) $\xi = 1$: $\tau(g) \sim g^{2}$ \hspace{0.5cm}(diffusive behavior).

Real dendrimers have energy fluctuations and quenched (slow) energy
disorder plays an important role in the energy transfer dynamics. We
have studied the effects of disorder on diffusion in the presence of a
linear potential by considering four models of disorder denoted by
$(i)$ - $(iv)$. These will be described below.

\subsection{Intergenerational Quenched Disorder}

Assuming that energy fluctuations in the segment lengths of the same
generation are identical, the diffusion is mapped into an effective
one-dimensional problem. We denote this type of disorder
intergenerational.

Intergenerational energy fluctuations can be introduced in two ways. In
the absence of correlations, we obtain the standard diagonal disorder
(random energy) model (i) where a random part $\epsilon_{n}$ is added
to the linear energy $\overline{\epsilon}(n)$. In the second form of
disorder (random force model(ii)), the energy differences $\Delta
\epsilon _{n}$ ($ = \epsilon(n+1) - \epsilon(n))$) are treated as random
variables with the positive average values $\overline{\epsilon}(n)$ to
ensure funneling. Both models are one dimensional and the MFPT is given
by eq.(10).

{\underline{ Model (i): Random intergenerational energy}}

In this model, all segments in a given
generation have the same random energy. Generational
energies are made random by adding a fluctuating part $\epsilon_{n}$
to the linear potential
$$
    \epsilon(n) = \overline{\epsilon}(n) + \epsilon_{n}.
$$ $\epsilon_{n}$'s are independent and identically distributed with
the average $\langle \epsilon_{n} \rangle =0$ and
$\overline{\epsilon}(n)$ is the linear potential defined in Eq.~(11). A
similar random energy model has been considered previously
\cite{murthy} for a discrete random walk.

Hereafter $\langle \rangle$ will denote the average over disorder. We
define $\eta_{n} = \exp(- \beta \epsilon_{n} )$ and assume $\langle
\eta ^{\pm 1} \rangle$ and $\langle \eta^{\pm 2} \rangle$ to be finite.
The subscript $n$ in $\langle \eta \rangle$ is omitted because the
random variable $\epsilon_{n}$'s are independent and identically
distributed. Hence

\begin{eqnarray}
 \xi_{n}\xi_{n-1}....\xi_{2}\xi_{1}
 & = &
 c exp[-\beta (\epsilon(n+1) -\epsilon(n))] \,
 c exp[-\beta (\epsilon(n) -\epsilon(n-1))] ....  \\
 & & \quad ... c exp[-\beta (\epsilon(3) -\epsilon(2))] \,
      c exp[-\beta (\epsilon(2) -\epsilon(1))] \nonumber \\
 & = &\xi_{0}^{n}  \{ exp [-\beta(\epsilon_{n+1} - \epsilon_{n})] \}.
\end{eqnarray}

Averaging over realizations of disorder gives

\begin{equation} \label{eq:}
  <\xi_{n}\xi_{n-1}....\xi_{2}\xi_{1}> =
   <\eta> <\eta^{-1}> \xi_{o}^{n}.
\end{equation}

Substituting this in Eq. (10) gives for the MFPT

\begin{equation} \label{eq:}
\langle \tau(g) \rangle \sim <\eta> <\eta^{-1}> \tau_{0}(g),
\end{equation}
where $\tau_{0}(g)$ is the MFPT (see Eq. (12)) for a linear potential
with no disorder.

Similar to the ordered case, the disorder averaged MFPT has three
distinct regimes: linear ($\xi_{0} < 1$), quadratic ($\xi_{0} = 1$) and
exponential ($ \xi_{0} > 1$) depending on the value of $\xi_{0}$. Even
though the $g$ dependence and the critical point ($\xi_{0} = 1$) of the
MFPT do not change by disorder, the magnitude of the disorder averaged
MFPT exceeds the corresponding value for the ordered system. Disorder
slows down the first passage diffusion. If the fluctuating part of the
energy $\epsilon_{n}$ has a Gaussian probability distribution
$P_{G}(\epsilon) = (1/ \sqrt{2 \pi} \lambda) exp \{-\epsilon^{2}/ 2
\lambda^{2} \} $, then

\begin{eqnarray}
  \langle \tau_{g} \rangle & \sim & <\eta> <\eta^{-1}> \tau_{0}(g)
 \nonumber \\
& = & \langle exp(-\beta \epsilon_{n}) \rangle
 \langle exp(\beta \epsilon_{n}) \rangle
  \tau_{0}(g) \nonumber \\
& = & exp(\beta^{2} \lambda^{2}) \tau_{0}(g).
\end{eqnarray}

The average MFPT is thus increased by the disorder-dependent factor of
$exp(\beta^{2} \lambda^{2})$.

So far we have considered only the average MFPT, but due to disorder
there will be fluctuations in the MFPT corresponding to various
realizations of disorder. The distribution of MFPT due to disorder may
be directly observed in single molecule spectroscopy \cite{26}. We have
calculated the rms fluctuations in the MFPT $<(\Delta \tau)^{2}> =
<(\tau - <\tau>)^{2}>$ in different regimes of $<\tau(g)>$.

In the linear regime ($\xi_{0} < 1$), only the terms linear in $\xi$ in
Eq. (10) are sufficient to yield the correct scaling (in terms of $g$)
for both $<\tau(g)>$ and $<\Delta \tau^{2}>$. Considering only the
linear terms in Eq.(10), we get

\begin{eqnarray}
 \tau (g) & \approx & \{ 1 \}
 + \{  \xi_{0} \, \exp [-\beta (\epsilon_{2} - \epsilon_{1})]+ 1 \}
 + \{  \xi_{0} \, \exp [-\beta (\epsilon_{3} - \epsilon_{2})]+ 1 \}
 + .... \nonumber \\
& & \quad   ....
 + \{  \xi_{0} \, \exp [-\beta (\epsilon_{g} - \epsilon_{g-1})] +1 \}
\nonumber \\
 & = & \xi_{0} \{ \exp (-\beta\epsilon_{2}) \, \exp (\beta\epsilon_{1})
+ \exp (-\beta\epsilon_{3}) \,  \exp (\beta\epsilon_{2})
+ ....  \nonumber \\
& & \quad + \exp (-\beta\epsilon_{g}) \, \exp (\beta\epsilon_{g-1}) \} + g.
\end{eqnarray}

Averaging over realizations of disorder gives

\begin{eqnarray}
 \langle \tau (g)  \rangle & = &
  (g-1) \, \xi_{0} \, \langle \eta \rangle \langle \eta^{-1} \rangle + g
%& \simeq & \frac{g} { 1 -
%\xi_{o} \langle \eta \rangle \langle \eta^{-1} \rangle}
\end{eqnarray}

To calculate the fluctuation of the MFPT around its average we need
$\langle \left( \tau(g) \right) ^{2} \rangle$. Taking the disorder
average of the square of the Equation

\begin{eqnarray}
 \langle \left( \tau(g) \right) ^{2} \rangle
& = &  (g-1) \, \xi_{0}^{2} \,
\langle \eta \rangle ^{2} \langle \eta^{-2} \rangle
+ 2(g-1) \, \xi_{0}^{2} \,
\langle \eta \rangle \langle \eta^{-1} \rangle \nonumber \\
& & \quad + 2 \left \{ {g-1 \choose 2} - (g-2) \right\} \,
\xi_{0}^{2} \, \langle \eta \rangle ^{2}
\langle \eta^{-1} \rangle ^{2} \nonumber \\
& & \qquad + 2g(g-1) \, \xi_{0} \,
\langle \eta \rangle \langle \eta^{-1} \rangle + g^{2}.
\end{eqnarray}

From the average MFPT eq. (19)

\begin{eqnarray}
\langle \left( \tau(g) \right) \rangle ^{2}
& = & (g-1)^{2} \, \xi_{0}^{2} \,
\langle \eta \rangle ^{2} \langle \eta^{-1} \rangle ^{2}
- 2g(g-1) \, \xi_{0} \, \langle \eta \rangle \langle \eta^{-1} \rangle + g^{2}.
\end{eqnarray}

Combining eqs. (20) and (21) we obtain for the fluctuation in the MFPT

\begin{eqnarray}
  \langle \left( \Delta \tau(g) \right) ^{2} \rangle
 & = & \langle \left( \tau(g) \right) ^{2} \rangle -
\langle \left( \tau(g) \right) \rangle ^{2} \nonumber \\
 & = & g \, \xi_{0}^{2} \, \left \{
\langle \eta^{2} \rangle \langle \eta^{-2} \rangle + 2 \langle \eta
\rangle \langle \eta^{-1} \rangle - 3 \langle \eta \rangle ^{2} \langle
\eta^{-1} \rangle ^{2} \right \}.
\end{eqnarray}

Hence the relative fluctuation in the linear regime ($\xi_{0} <1$)
will be given by

\begin{eqnarray}
\frac{\Delta \tau}{\tau} & = &
\frac{ \langle (\Delta \tau (g))^{2} \rangle ^{1/2}}
{\langle \tau(g) \rangle}
\nonumber \\
& \simeq & \frac{ \sqrt{g} \, \xi_{0} \, \left \{
\langle \eta^{2} \rangle \langle \eta^{-2} \rangle
+ 2 \langle \eta \rangle \langle \eta^{-1} \rangle
-3 \langle \eta \rangle ^{2} \langle \eta^{-1} \rangle ^{2} \right \}^{1/2}}
{g \{ 1 + \xi_{o} \langle \eta \rangle \langle \eta^{-1} \rangle \}}
\end{eqnarray}

At the transition point $\xi_{0} = 1$, we have to keep all terms in Eq.
(10) in the calculation of $\Delta \tau / \tau$. Some tedious
calculations finally yield $\delta \tau / \tau \simeq 1/ \sqrt{g}$, the
same scaling as found above in the linear regime.

In the
exponential regime ($\xi_{0} > 1$), the MFPT is dominated by the
single term in expression (10). In this case
$$
   \tau(g) \simeq \xi_{g-1} \xi_{g-2}....\xi_{2} \xi_{1}
$$

The average MFPT is then $ <\tau(g)> \simeq \xi^{g}_0 <\eta ><\eta^{-1}
>$ for large $g$, as expected. In this approximation $
<(\tau(g))^{2}> \simeq  \xi^{2g}_0 <\eta^2 ><\eta^{-2}
>$, and the fluctuation in the MFPT is

\begin{eqnarray}
    \frac{\Delta \tau}{\tau} & \simeq [&\frac{ <\eta^2 ><\eta^{-2}>
    }{(<\eta ><\eta^{-1})^2}-1]^{\frac{1}{2}}
\end{eqnarray}

which is a constant independent of g.

{\underline{Model (ii): Random intergenerational force}}

In this model energy-differences between consecutive generations are
randomly distributed, but again all segments in the same generation are
identical. Fluctuations of the energy levels are correlated and the
energy differences $\Delta \epsilon_{n}$ are assumed to be identically
distributed according to some probability distribution $P(\Delta
\epsilon_{n})$. This diffusion model was studied extensively in various
contexts and was reviewed in \cite{bouchaud}. The effect of disorder is
more pronounced in this case compared with the random energy model. The
average MFPT is given by
\begin{equation} \label{eq:}
 <\tau(g)>  = <\xi> \frac{<\xi>^{g} -1}{(<\xi> -1)^2} -
                          \frac{g}{<\xi> -1}
\end{equation}
$<\tau(g)>$ again has three regimes, but the transition occurs at
$<\xi> = 1$. Note that in model (i), this condition was $\xi_{0} = 1$,
and disorder did not change the transition point. In the present model,
however, the transition point is shifted due to the stronger effect of
disorder. In some regimes of the MFPT, the effect of disorder is so
strong that the typical diffusion time differs from the MFPT. This
indicates a broad distribution of the MFPT where the average is
affected by the rare configurations with long first passage time.  The
more representative ``typical'' diffusion time is calculated as
$\tau_{typ} = exp \langle \log \tau (g) \rangle$. Following
\cite{bouchaud}, we define different scaling regimes of the MFPT and
$\tau_{typ}$:

(a) For $<\xi>\hspace{0.1cm} < 1$, $<\tau(g)>$ is linear in $g$.
$<\xi^{2}>$ determines the fluctuation around the average.
$<\tau^{2}(g)> \sim g^{2}$ as long as $<\xi^{2}> < 1$ and hence the
relative fluctuation $\delta \tau / \tau$ goes to a constant (similar
to the random energy case). For $<\xi^{2}> > 1$, however,
$<\tau^{2}(g)>$ grows exponentially with $g$ as the typical behavior
starts to differ from the MFPT.

If $<\xi>$ $> 1$, $<\tau(g)>$ is always exponential in $g$. However,
$\tau_{typ}$ varies depending on the value of $ <ln \xi> $ (whether
$\langle ,=, or \rangle 1$).

(b) $<\xi>$ $> 1$, $ <ln \xi>$ $< 0$: The typical first passage time ($
\tau_{typ}(g) = exp<log \tau(g)> $) does not follow the average
exponential behavior, but rather grows like a power law $g^{\alpha}$
with $\alpha = <\delta (ln \xi)^{2}> / 2 <ln \xi>$. For a Gaussian
probability distribution $P_{G}(\Delta \epsilon)$ the above formula
yields $\alpha = \beta ^{2} \lambda ^{2} / 2(ln \xi_{0})$.

(c) $<\xi>$ $> 1$, $ <ln \xi> = 0$ \cite{sinai}: At this transition
point (the ``Sinai point''), $<\tau(g)>$ is exponential in $g$, but
the more representative $ \tau_{typ} (g) $ is only exponential
in $ \sqrt{g}$ \cite{gold}.

(d) $<\xi>$ $> 1$, $ <ln \xi>$ $> 0$: In this regime, both $<\tau(g)>$
and $ \tau_{typ} (g)$ diverge exponentially with $g$.

\subsection{Intersegment Quenched Disorder}

When energy fluctuations exist within the same generation (intersegment
disorder), the system can no longer be mapped into a one dimensional
model and we have to use eqs. (4) and (6) to compute the MFPT for the
actual tree structure. The energy of site $q_{n}$ is denoted
$\epsilon_{q}(n)$. The random variables $\xi_{n}$'s become matrices
$\xi^{p,q}_{n}$ where ($p,q$) represents $q$th point of $n$th and the
$p$th point of the ($n+1$)th generation. (note that $c$ is not included
in this definition). The detailed balance ratio becomes $\xi^{p,q}_{n}
= \exp \{ - \beta \Delta \epsilon_{pq}(n) \} = \xi \, \ exp \{ -\beta
(\epsilon_{p_{n+1}} - \epsilon_{q_{n}})\}$ , where $\epsilon_{q_{n}}$
is the fluctuating part of the energy and $\xi = \exp \{- \beta
(\overline{\epsilon}(n+1) - (\overline{\epsilon}(n)) \}$ (so $\xi_{0} =
c \xi$). From now on we assume the branching ratio $c = 2$. The MFPT
for a particle released at the peripheral site $1_{g}$ of the
g-generational tree is

\begin{eqnarray}
  \langle \tau^{1_{g}}(g) \rangle  & = & \{1\}  \nonumber \\
  & + & \{ (\xi^{1,1}_{1} + \xi^{1,2}_{1}) + 1 \} \nonumber \\
  & + & \{ \xi^{1,1}_{1}(\xi^{1,1}_{2} + \xi^{1,2}_{2})
 + \xi^{1,1}_{1}(\xi^{2,3}_{2} + \xi^{2,4}_{2})
 + (\xi^{1,1}_{2} + \xi^{1,2}_{2}) + 1 \} \nonumber \\
& + &....... \nonumber \\
& + & \Big \{ \xi^{1,1}_{1} \left(\xi^{1,1}_{2}\Big(....()....\Big)
                  + \xi^{1,2}_{2}\Big(.....()....\Big) \right)
 + \xi^{1,2}_{1} \left(\xi^{2,3}_{2}\Big(....()....\Big)
                  + \xi^{2,4}_{2}\Big(.....()....\Big) \right) + ..
\nonumber \\
 & & \qquad  ........
 + (\xi^{1,1}_{g-1} + \xi^{1,2}_{g-1})  + 1 \Big \}
\end{eqnarray}

We define the initial site average MFPT (ISA-MFPT) as:
\begin{equation}
\overline{\tau (g)} = \frac{1}{N_{g}} \sum_{i_{g}=1}^{N_{g}}
\tau^{i_{g}(g)},
\end{equation}

where $N_{g} = (1+c)c^{g-1}$ is the number of peripheral sites (leaves
of the tree).

$\overline{\tau (g)}$ may be expressed again (see Eq. (9)) as a sum over
ISA-MRT:
\begin{equation}
\overline{\tau (g)} = \sum_{n=1}^{g} \overline{t_{n}}.
\end{equation}

Each of the  $\overline{t_{n}}$ is made of sum of products over $\xi$'s
up to generation $n$. Formally it may be expressed as

\begin{equation}
\overline{t_{n}} = 1 +  \sum_{k=1}^{n-1} \{ \frac{1}{c^{k-1}}
                 \sum_{i_{k}=1}^{c^{k-1}} T^{n}_{i_{k}} \} ,
\end{equation}
where $T^{n}_{i_{k}}$ stands for the contribution to $\overline{t_{n}}$
from a subtree rooted at site $i_{k}$ on generation $k$ and spreading
out to the $n^{th}$ generation. This contribution is made of products
of $(n-k)$ $\xi$'s along the $c^{n-k}$ paths going from the site
$i_{k}$ to the $c^{n-k}$ different sites on generation  $n$. The $k$'th
generation has $c^{k-1}$ rooted subtrees with $i_{k} = 1, ..... ,
c^{k-1}$. The contributions from subtrees rooted at the same generation
are independent from each other. Symbolically,

\begin{eqnarray}
    T^{n}_{i_{k}} & = & \sum_{\Gamma _{m}}
     \left( \prod_{{pq}_{m} \in \Gamma _{m}} \xi^{p,q}_{m} \right)
\nonumber \\
 & = & \sum_{\Gamma _{m}} \exp \left(- \beta \sum_{{pq}_{m}}
                    \Delta \epsilon _{pq}(m) \right),
\end{eqnarray}

where $\Gamma_{m}$ are all paths on this subtree from $i_{k}$
to generation $n$.

{\underline{Model (iii): Random intersegment energy}}

In a fully disordered tree all segment energies ($ \epsilon
_{i_{n}}\vspace{0.1cm}$)are random and uncorrelated . The techniques
applied in model (i) can be simply extended to this case with almost
identical results (Fig. 2 shows the average MFPT for both the models).

In the linear regime, the MFPT expression can be approximated by
neglecting terms $\xi^{2}$ and higher. For $c = 2$ we have

\begin{eqnarray}
 \tau ^{1_{g}}(g) & \approx & \{ 1 \}  \nonumber  \\
 & + &  \left \{  \xi \, \exp [-\beta(\epsilon_{1_{2}} - \epsilon_{1_{1}})]
  + \xi \, \exp [-\beta(\epsilon_{1_{2}} - \epsilon_{1_{1}})] + 1 \right \}
\nonumber \\
 & + &  \left \{  \xi \, \exp [-\beta(\epsilon_{1_{3}} - \epsilon_{1_{2}})]
  + \xi \, \exp [-\beta(\epsilon_{2_{3}} - \epsilon_{1_{2}})] + 1 \right \}
\nonumber \\
 & + & ....... \nonumber \\
 & + &  \left \{  \xi \, \exp [-\beta(\epsilon_{1_{g}} - \epsilon_{1_{g-1}})]
  + \xi \, \exp [-\beta(\epsilon_{2_{g}} - \epsilon_{1_{g-1}})] + 1 \right \}
\nonumber \\
 & = & \xi \{ \exp [-\beta(\epsilon_{1_{2}} - \epsilon_{1_{1}})]
+ \exp [-\beta(\epsilon_{1_{2}} - \epsilon_{1_{1}})]
+ \exp [-\beta(\epsilon_{1_{3}} - \epsilon_{1_{2}})]
+ \exp [-\beta(\epsilon_{2_{3}} - \epsilon_{1_{2}})]
\nonumber \\
& & \quad + .....
+ \exp [-\beta(\epsilon_{1_{g}} - \epsilon_{1_{g-1}})]
+ \exp [-\beta(\epsilon_{2_{g}} - \epsilon_{1_{g-1}})] \} + g.
\end{eqnarray}

Upon averaging over realizations of disorder we obtain

\begin{eqnarray}
 \langle \tau ^{1_{g}}(g)  \rangle & = &
 2 (g-1) \, \xi \, \langle \eta \rangle \langle \eta^{-1} \rangle + g
\nonumber \\
 & = &
 (g-1) \, \xi_{0} \, \langle \eta \rangle \langle \eta^{-1} \rangle + g
\quad \textrm{for g large},
\end{eqnarray}
where $\xi_{0} = 2 \xi$ with $c=2$.

The average MFPT is identical for models (i) and (iii). The fluctuation
of the MFPT around its average are connected to $\langle \left( \tau(g)
\right) ^{2} \rangle$. Taking the disorder average of the square of the
Eq. (31)

\begin{eqnarray}
 \langle \left( \tau(g) \right) ^{2} \rangle
& = &  2(g-1) \, \xi^{2} \,
\langle \eta^{2} \rangle \langle \eta^{-2} \rangle
+ 2(g-1) \, \xi^{2} \,
\langle \eta \rangle ^{2} \langle \eta^{-2} \rangle
+ 2 \{2(g-1)\} \, \xi^{2} \,
\langle \eta \rangle \langle \eta^{-1} \rangle \nonumber \\
& & \quad + 2 \left \{ {2(g-1) \choose 2} - (g-1) - 2(g-2) \right\} \,
\xi^{2} \, \langle \eta \rangle ^{2}
\langle \eta^{-1} \rangle ^{2} \nonumber \\
& & \qquad + 2\{2g(g-1)\} \, \xi \,
\langle \eta \rangle \langle \eta^{-1} \rangle + g^{2}.
\end{eqnarray}

The average MFPT eq. (32) yields

\begin{eqnarray}
\langle \left( \tau(g) \right) \rangle ^{2}
& = & 4(g-1)^{2} \, \xi^{2} \,
\langle \eta \rangle ^{2} \langle \eta^{-1} \rangle ^{2}
- 4g(g-1) \, \xi \, \langle \eta \rangle \langle \eta^{-1} \rangle + g^{2}
\nonumber \\
& = &  (g-1)^{2} \, \xi_{0}^{2} \,
\langle \eta \rangle ^{2} \langle \eta^{-1} \rangle ^{2}
- 2g(g-1) \, \xi_{0} \, \langle \eta \rangle \langle \eta^{-1} \rangle
+ g^{2}
\end{eqnarray}

Eqs. (33) and (34) yield for the fluctuation in the MFPT

\begin{eqnarray}
  \langle \left( \Delta \tau(g) \right) ^{2} \rangle
 & = & \langle \left( \tau(g) \right) ^{2} \rangle -
\langle \left( \tau(g) \right) \rangle ^{2} \nonumber \\
 & = & g \, \xi_{0}^{2} \, \left \{
\frac{1}{2} \langle \eta^{2} \rangle \langle \eta^{-2} \rangle +
\frac{1}{2} \langle \eta \rangle ^{2} \langle \eta^{-2} \rangle +
\langle \eta \rangle \langle \eta^{-1} \rangle - 2 \langle \eta \rangle
^{2} \langle \eta^{-1} \rangle ^{2} \right \}
\end{eqnarray}

The relative fluctuation ${\Delta \tau}/{\tau}$ $\sim {1}/{\sqrt g}$.
Scales with $g$ the same way as in the case of intergenerational
disorder, but the amplitude is smaller in this model. This can be seen
by comparing Eq. (22) and Eq. (35) and since $\langle \eta \rangle ^{2}
\le \langle \eta^{2} \rangle$. This is also verified by our numerical
calculation (Fig. (3)) where we kept all the terms in the MFPT
expression. In the exponential regime, the relative fluctuations
${\Delta \tau}/{\tau}$ will still saturate to a $g$-independent value,
but the latter will be also smaller than that reached with Model (i).

The differences between models (i) and (iii) may also
stem from the additional
fluctuations in the MFPT for the latter model which arise from distinct
initial sites at the periphery. For initial excitations starting at two
different peripheral sites (on the same dendrimer) separated by an
ultrametric distance $p$ the fluctuations in the MFPT is given by $$ <
(\Delta \tau(g))^{2} > \sim p. $$ Averaging over all possible
peripheral sites, the fluctuation scales with $g$ the same way as the
dendrimer-to-dendrimer fluctuations discussed above. The amplitude of
the fluctuation is different from that of the dendrimer-to-dendrimer
fluctuations and depends heavily on the branching ratio $c$ of the
tree.

{\underline{ Model (iv): Random intersegment force}}

This model assumes that all energy-differences ($ \Delta
\epsilon_{pq}(n) \equiv \epsilon_{p}(n+1) - \epsilon_{q}(n)$) between
neighboring segments are random. We observe a new transition apart from
the usual linear to exponential regime transition of the MFPT. This is
a one-step replica symmetry breaking transition from a weakly
disordered (high-temperature) phase to highly disordered
(low-temperature) phase. In the weakly disordered phase all paths
contribute to $\overline{\tau(g)}$, whereas in the highly-disordered
phase $\overline{\tau(g)}$ is dominated by a few paths. Similar dynamic
transition was predicted \cite{wolynes} for kinetic pathways in protein
folding.

This transition is occurring only in this model.  It cannot occur in
models (i) and (ii) since in these models there is effectively a single
pathway from the periphery to the center (i.e. all pathways are
equivalent).  It cannot occur in model (iii) since the random-energy
effect is very limited.  For any path, independent of its length, only
the random energies at the initial and final sites are contributing.
This is insufficient to cause an energetic disparity which will
overcome the entropic advantage of having a maximum number of path
contributing.

In the low-temperature phase the distribution of $ \xi^{p,q}_{n} =
\exp(- \beta \Delta \epsilon _{p,q}(n))$ becomes very broad. The
important contributions will come from those paths for which
$\sum_{(pq)_{n}} \Delta \epsilon _{p,q}(n)$ is large and negative. We
also assume that the constant force
$\overline {\epsilon} = \epsilon_{o}$ is weak and the
system is in the exponential regime. The linear regime will be
discussed later. We thus focus on the large $\xi$ regime (it is
sufficient for them to be typically larger than $1/c$). In this regime,
the largest contributions will come from the longest paths, and if a
qualitative change in the behavior will occur it will be noticeable
first in the dominant contribution from the largest of all subtrees. We
will thus look for a new behavior in the ISA-MRT by examining at the
maximal tree rooted at  $k=1$. The same argument may be repeated while
looking at $\overline{\tau(g)}$: its dominant contribution in this
regime will come from $\overline{t_{g}}$ namely the ISA-MRT on the most
extremal generation with $n=g$.

Searching for a possible abrupt change in the properties of ISA-MFPT,
we will therefore examine the largest tree contributing to
$\overline{t_{g}} \sim T_{1}^{g}$ since it has the largest energy
disparity. As long as its behavior is normal, so will be all other terms.
Once its behavior changes, it will affect that of $\overline{t_{g}}$
and then that of $\overline{\tau (g)}$, to which it makes the largest
contribution.

The contribution of $T_{1}^{g}$ is akin to a partition function of a
so-called ``random-directed polymer'' on a Cayley tree which is known to
have a glass transition (in the thermodynamic limit). Physically the
transition is between a high temperature phase $T \ge T_{g}$ at which
all paths contribute, while for $T \le T_{g}$ only a finite number of
them do. The transition is of one-step replica symmetry breaking (1RSB)
type. The $n \rightarrow 0$ replica-trick was used to study its
transition point. A complete analysis using this trick is available
\cite{derrida}. Here we present a simpler, more intuitive, outline. In
the high temperature regime $\beta < \beta_{g}$ it may be shown that
all moments of the MFPT (or the ``partition function'' , in the polymer
picture) obey $\langle T^{k} \rangle = \langle T \rangle ^{k}$. Hence
all thermodynamic properties of the quenched system are given by those
in which the disorder is annealed. In this regime $<T_{1}^{g}>$ is
easily computed and found to be
\begin{equation} \label{eq:}
 <T^{g}_{1}> = (c<\exp (-\beta \Delta \epsilon)>) ^{g}.
\end{equation}
For large $g$ we have

\begin{equation} \label{eq:}
 <T^{g}_{1}> = [\exp \{ - \beta f \}]^{g},
\end{equation}
where $ f(\beta) = - \frac{1}{\beta} \ln  <T^{1}_{g}> $ is the
free energy per generation.

When the random energy difference $\Delta \epsilon$ follows a Gaussian
distribution $P_{G}(\Delta \epsilon) = (1/ \sqrt{2 \pi} \lambda) exp
\{-\frac{(\Delta \epsilon - \epsilon _{0})^{2}}{ 2 \lambda ^{2}} \}$,
the free energy is given by $$ \beta f = \beta \epsilon_{0} - \ln c -
\frac{1}{2} \beta ^{2} \lambda ^{2}. $$ The transition point can be
determined by the vanishing point of the entropy $S (= - \frac{\partial
f}{\partial T} = \ln c - \lambda^{2}/2 T^{2}$ , where we assumed $K_{B}
= 1$). This yields $T_{g} = \lambda / \sqrt{2 \log c}$, which is the
transition temperature from the weakly disordered to the highly
disordered phase.

For low temperatures $\beta > \beta_{g}$ the entropy remains zero. This
is achieved \cite{derrida} by expressing the free energy as $f =
\epsilon_{0} - \frac{\ln c}{m \beta} - \frac{m \beta \lambda^{2}}{2}$,
and $m \in [0,1]$ is chosen such that $f$ is maximal. For $\beta <
\beta_{g}$, $m=1$ and the previous result holds. For $\beta >
\beta_{g}$, $ \frac{\partial f} {\partial m}=0$ (which implies
$\frac{\partial f} {\partial T}=0$) yields

\begin{eqnarray}
m = \frac{\sqrt{2 \ln c}}{\beta \lambda}
= \frac{\beta_{g}}{\beta} < 1.
\end{eqnarray}

For a general distribution the requirement is to maximize

$$ f(m) = - \frac{1}{m \beta} \left (\ln c +
\ln \langle \exp (- m \beta \Delta \epsilon \rangle) \right).$$

The physical meaning of $m$ is explained by the existence of a finite
(in the thermodynamic $g \rightarrow \infty$ limit) number of paths
that contribute to $T_{1}^{g}$, and therefore to $\overline{t_{g}}$ and
$\overline{\tau (g)}$ as well. To see that, we need to define the
overlap $q$ between the contribution of two different products (paths)
in the sum (and its distribution $P(q)$) which plays here the role of
an order parameter. It is defined as the weighted (over all paths)
fraction of the way the two paths will go together on the same
segments, averaged over the disorder. At high temperatures all paths
are essentially equivalent and the probability for any overlap
vanishes. At $T=0$ there is a single dominant path with the minimal
energy (the ``ground state'') that makes the only contribution and thus
both paths will overlap all the way giving $q=1$ with probability $1$.
It was shown \cite{derrida} that for this system the only two possible
overlaps are $q=0$ and $q=1$. Their respective probabilities, however
changes in the glassy phase and is given by:

\begin{eqnarray}
 P(q) & = & \delta (q)  \hspace{2in} \textrm{for $\beta < \beta_{g}$
 and} \nonumber \\
 P(q) & = &  m \delta (q) + (1-m) \delta (q-1) \hspace{1in}
  \textrm{for $\beta > \beta_{g}$}
\end{eqnarray}

where $m = \frac{T}{T_{g}}$ was introduced before.

So far the discussion was limited to the exponential regime. Increasing
$\epsilon_{o}$ will lead to the transition into a linear regime at
$\epsilon _{o} = \epsilon^{*}_{o} (\beta)$. For $\beta < \beta_{g}$ the
transition is given by the expected ``annealed'' condition $c <\xi> =
1$ which yields $\epsilon^{*}_{o} = \frac{\ln c}{\beta} + \frac{1}{2}
\beta \lambda^{2}$. For $\beta > \beta_{g}$, in the glassy phase
free-energy is given by

\begin{equation} \label{eq:}
 f(m,\beta) = \frac{\ln c}{m \beta} + \frac{m \beta \lambda^{2}}{2}.
\end{equation}

Since $m \beta = \beta_{g}$ the critical force remains constant within the
glassy phase
for $T_{g} < T < 0$ and is equal to its value at the transition point
($\beta = \beta_{g}$), which is given by

\begin{equation} \label{eq:}
 \epsilon_{0}^{*} (\beta > \beta_{g})
=\frac{\ln c \lambda}{\sqrt{2 \ln c}} + \frac{\sqrt{2 \ln c} \lambda}{2}
= \lambda \sqrt{2 \ln c}.
\end{equation}

The expected ($\epsilon_{o}, T$) phase diagram is shown in Fig. (4)

We should also note that crossing $\beta_{g}$ within the linear regime
yields only a glassy crossover and not a sharp transition. This is due
to the fact that the ISA-MRT converges to a finite value even as $g
\rightarrow \infty$. The longer paths thus make an exponentially small
(in their length) contribution, the system is dominated by paths of
finite length and the ``glass transition'' takes place in an
effectively finite system.

\section{MFPT FOR THE NON-LINEAR TDHF POTENTIAL}

In the previous section the funneling energy was assumed to vary
linearly with the generation number $n$ ($=0,1,2...g$). However
electronic structure calculations by Tretiak, Chernyak, and Mukamel
\cite{TCM1} show a strong nonlinear dependence of the exciton energy
(on $n$).

The funneling effect of extended dendrimers originates from the
variation of the segments length.  We denote by $n (=1,2, \ldots, g$)
the $n^{th}$ generation starting from the center. The peripheral $n=g$
generation is made from $m=1$ monomer. The number of monomers increases
by one going from one generation to the next towards to the center. The
number of acetylene monomers in the $n^{th}$ generation is thus $l = g
- (n - 1)$. The $n$ dependence of the excitation energy was computed
using time-dependent Hartree Fock (TDHF) technique and fitted to the
form \cite{TCM1}

\begin{equation} \label{eq:}
\epsilon(n) = A(1+\frac{L}{g-(n-1)})^{0.5}
\end{equation}

with $A=2.80 \pm 0.02 (e.v.)$ and $L=0.669 \pm 0.034$.

\subsection{Ordered Dendrimer}

Using eq. (42), the detailed balance ratio becomes

\begin{equation}
 \xi_{n} = c \exp{- \frac{A}{kT}\{ (1+ \frac{L}{g-n)})^{0.5} -
                   (1+ \frac{L}{g-n+1})^{0.5} \}}
\end{equation}

The points of the same generation are identical and we can calculate
the MRT and MFPT for this potential using the expressions (8) and (10)
for one-dimensional random walk starting at the periphery. The mean
time spent by an exciton in the nth generation is given by

\begin{equation} \label{eq:}
 t_{n} = \exp[{-A(1+\frac{L}{g-(n-1)})^{0.5}}] \sum_{r=1}^{n}
             c^{(n-1)} \exp[{A(1+ \frac{L}{g-(r-1)})^{0.5}}]
\end{equation}
The total time to reach the trap will be sum of all the $\tau_{n}$'s,
and the MFPT is
\begin{eqnarray}
 \tau(g) = \sum_{n=1}^{g} \tau_{n}
                  = \sum_{n=1}^{g} \exp{-A(1+\frac{L}{g-(n-1)})^{0.5}}
                   \sum_{r=1}^{n} c^{(n-1)} \exp{A(1+ \frac{L}{g-(r-1)})^{0.5}}
\end{eqnarray}

Fig. 5 shows that the MFPT for this potential is quite different
compared to that for the constant potential difference. We find that it depends
linearly on g for the first few generations, but gradually changes to
exponential with increasing $g$.

  The behavior of the MFPT can be better understood in terms of MRT,
which is not a monotonic function of n (see Fig. 6) for the TDHF
potential. At the generations near the periphery ($g - n \ll g$) the
energy difference between generations is large and the funneling force
is so strong that it overcomes the outward entropic force of $c \ge
2$. Near the center, on the other hand, the larger is $g$, the weaker
is the funneling force and the entropic term dominates. The competing
effects nearly cancel at some intermediate generation $n = n^{*}(g)$
where the exciton spends maximum time. For large $g$, an estimate for
$n^{*}$ may obtained from the recursion relation satisfied (from Eq. (8))
by the MRT's

\begin{equation} \label{eq:}
                   t_{n} = \xi_{n-1} t_{n-1} + 1
\end{equation}
For $t_{N}$ to be maximal, $ {t_{n+1}}/{t_{n}} = \xi_{n} + {1}/{t_{n}}
< 1$, hence $\xi_{n} < 1$. Similarly requiring $
{\tau_{n}}/{\tau_{n-1}} > 1$ yields $\xi_{n-1} + {1}/{\tau_{n-1}} > 1$,
so for large $\tau_{n-1}$ (or equivalently large g) we may take
$\xi_{n-1} \geq 1$. $\tau_{n}$ is therefore maximized as $\xi_{n}
\rightarrow 1$ from below. Using expression (43) we obtain $n \sim (g -
\kappa + 1)$ where $\kappa (=\kappa(A, L, c))$ is a constant
(independent of $n$ and $g$) and $n^{*}(g)$ is the closest integer.
This formula works well for our system for $g \ge 12$ and holds
approximately (off by only one generation)
for $ 7 \le g < 11$.
Clearly, the larger the $g$, the better is the estimate of
$n^{*}(g)$. The reduced free energy $u(n) = \beta \epsilon (n) - n \log
c$ is plotted in Fig. 7. The MRT is maximal at the site at which the
free energy is minimum. For large $g$, both $n^{*}(g)$ and the energy
difference between the center and $n^{*}$ increases with $g$.
As a result, the time to reach the center from $n^{*}$ grows
exponentially with $g$ (the time to arrive at $n^{*}$ from $n = g$ is
always much shorter).If g is small enough, however, the value of
$n^{*}$ and $u(n^{*})$ increase only weakly  with $g$ and the MFPT
dependence on $g$ may be approximated by a power series (dominated
initially by the linear term).

For fixed g, the MFPT may be changed by variations of the coordination
number (c) and the temperature (T). Decreasing the temperature makes
the funneling more effective. As a result, the crossover from the
linear to the exponential regime occurs at a higher $n$. Increasing c
has a similar effect: both the outbound ``entropic force'' which
competes with the energy funneling and the MFPT increase (Fig. 5).

\subsection{Random Energy Fluctuations}

We introduced disorder by adding a fluctuating part to the energy from
an uniform probability distribution, similar to the random energy
models $(i)$ and $(iii)$ for the linear potential. The maximum value
of the fluctuation is taken to be $\pm$ $4\%$ of the energy value.
Intergenerational and intersegment types of
disorder were considered separately. For the former case we used
the MFPT expression (10) obtained for one-dimensional random walk in TDHF
potential. For the latter, the MFPT formula (26) for
a Cayley-tree can be used to calculate the time spent by an exciton in
the n'th generation starting from a point (the points are labeled as
$1,2,3... 2^{g-1}$) at the periphery is given by

\begin{equation} \label{eq:}
t_{n} = \sum_{m=1}^{n} \{ \exp[{\beta \epsilon(m,1)} ]
           \sum_{r=1}^{n-m} \exp[{- \beta \epsilon(n,r)}] \},
\end{equation}
where $\epsilon(n,r)$ is energy of the rth branch of the nth generation
of the tree. r can take values $1,2,3.....2^{n-1}$. The MFPT is
obtained by summing all the $t_{n}$'s. The effect of disorder was
studied by numerically averaging the MFPT over different realizations
of the disorder. Fig. 8 shows disorder averaged ($\sim 10^{4}$
realizations) MFPT for both intergenerational and intersegment
disorders and also the case with no fluctuation. Disorder seems to work
against the funneling and increases the MFPT. The effect of disorder is
more pronounced in the exponential regime of the MFPT (Fig. 8). As for
previously discussed disordered linear potential the disorder-averaged
MFPT is similar for intergenerational and intersegment disorder,
consistent with our analytical calculation for models $(i)$ and
$(iii)$. However the fluctuations around the average MFPT $\delta
\tau(g)$, are different (Fig. 8) especially for large $g$ and the
relative fluctuation saturates to a larger value for intergenerational
disorder.

\section{GEOMETRIC DISORDER: DIFFUSION ON A TREE WITH RANDOM
CONNECTIONS}

Different forms of generic networks have drawn much recent interest.
The small world model was introduced by Watts and Strogatz \cite{watts}
and received a considerable attention  in the past two
years\cite{newman}. The model assumes a regular lattice and fixed
number of nearest neighbor connections with a few randomly chosen
connections between vertices. The essential features of this model are:
(i) High local connectivity which resembles a regular graph. (ii) The
average distance between any two points scales as $ \sim \log L$, where
$L$ is the linear size of the graph. This is an important property of a
random graph.

Inspired by this model we have studied the effect of a few random
connections on the exciton diffusion. We first considered the MFPT for
a one-dimensional chain with some random connections made between any
two vertices with a fixed small probability. For a given chain length
$L$, there exists a critical probability for random connections ($
p_{c} \sim L^{1/d} $ \cite{watts}), above which the small world effects
are observed. We have performed a numerical simulation of the MFPT for
a one-dimensional small world chain using a generalized form of the
master equation 1., where we allow a few hopping between vertices
chosen randomly. We then averaged over different realizations of random
connections. When $ p > p_{c}$, we found that the average MFPT scales
logarithmically (instead of linearly)with $g$ (Fig. 9) for $\xi_{0}$ $<
1$ . When $\xi_{0}$ $> 1$, the disorder averaged MFPT grows almost
linearly instead of exponentially with $g$ (Fig. 9). This may be
rationalized in terms of the average distance between any two vertices
which is supposed to behave as $\log g$ , where $g$ is the length of
the system. Hence we can approximate a small world linear chain of
length $g$ by a regular linear chain of length $\log g$. Then the
average MFPT in a small world 1-dimensional chain will be linear in
$\log g$ ({\it i.e.} $\tau(g) \sim \log g$) for $\xi_{0}$ $< 1$. For
$\xi_{0}$ $> 1$, the MFPT is exponential in $\log g$ and hence $\tau(g)
\sim \exp (\alpha \log g) = g^{\alpha}$, where $\alpha$ is a constant.
In our simulations $\alpha \sim 1$, so the MFPT grows linearly. As the
probability $p$ of random connections is decreased, the onset of the
small world effect occurs at larger system sizes.

On a tree we add a few random connections between vertices of different
generations. The resulting diffusion on such a ``small world tree''
yields results similar to the one dimensional case. Disorder averaged
MFPT again show logarithmic ($\xi_{0}$ $< 1$) or power-law ($\xi_{0}$
$> 1$) scaling with the number of generation $g$.

\section{CONCLUSIONS}

We have investigated analytically the diffusion on a disordered
Cayley tree in the presence of a linear potential. Fluctuations around
the MFPT showed different forms of scaling (in terms of $g$) depending
on the types of disorder and the scaling regime of MFPT. For a specific
form of disorder (random intersegment force) we found a new dynamic
transition in the MFPT. In the low temperature (highly disordered)
phase, the MFPT is dominated by a few paths. This transition resembles
the one-step replica symmetry breaking glass transition found in other
disordered systems like random directed polymers.

We have also considered exciton diffusion on dendrimers
with a nonlinear funneling energy. For dendrimers with a
nonlinear potential larger than a specific size, even though the number
of light-absorbing sites at the periphery increases, the MFPT starts to
grow exponentially (with $g$), resulting in slow exciton trapping.
Quenched disorder slows down the exciton diffusion towards the center.
This effect is more pronounced when the MFPT of the corresponding
ordered system is in the exponential regime. Hence to achieve an
efficient funneling of excitons, the number of generations must be
restricted to some optimal value. We have determined this optimal size
for a particular class of phenylacetylene dendrimers. For large
dendrimers, the free energy attains its minimum at $n^ {\star} (g)$
where an exciton spends most of its time during its journey towards the
center. Assuming a steady supply of long-lived photoexcitations, the
excitons will start gathering at $n^{\star} (g)$, rendering the
single-exciton picture invalid. In that case, we have to consider
exciton-exciton interaction and annihilation processes[25]

Finally, we have considered diffusion on a lattice with a few random
connections (``small world'') between any two vertices. The effect of
random connections is to reduce the effective system size
logarithmically. Numerical calculations show that the disorder averaged
MFPT is also decreased by a logarithmic factor. Hence there will be an
exponential gain in the exciton trapping. This might have interesting
consequences for information propagation through hierarchical social
communication networks.

\section{acknowledgments}
This work was supported by the Chemical Sciences Division of the Office
of Basic Energy Sciences of DOE.

\appendix{\bf Appendix}

The MFPT Eq. (10) can be expressed as

\begin{eqnarray}\label{eq:}
 \tau(g) & = & \{ 1 \}  \nonumber \\
 & + & \{ \xi_{1} + 1 \}  \nonumber \\
 & + & \{ \xi_{2} \xi_{1} + \xi_{2} + 1 \}  \nonumber \\
 & + &................   \nonumber \\
 & + & \{ \xi_{n}\xi_{n-1}....\xi_{2}\xi_{1}
    + \xi_{n}\xi_{n-1}....\xi_{2} + ...... \xi_{n}\xi_{n-1} +
    \xi_{n} +1  \}  \nonumber \\
 & + & .................  \nonumber \\
 & + & \{ \xi_{g-1}\xi_{g-2}....\xi_{2}\xi_{1}
  +  \xi_{g-1}\xi_{g-2}....\xi_{2} + ...... \xi_{g-1}\xi_{g-2} +
    \xi_{g-1} +1  \}
\end{eqnarray}

where $T_{i} = 1$ is assumed.

For the linear potential the sum in Eq. (10) assumes the form

\begin{eqnarray}
\tau(g) & = & \{1\}  \nonumber \\
        & + & \{\xi + 1 \}  \nonumber \\
        & + & \{\xi^{2} + \xi + 1\}  \nonumber \\
        & + & ..........   \nonumber \\
        & + & \{\xi^{n} + \xi^{n-1} + ....+\xi^{2} + \xi + 1\} \nonumber \\
        & + &...........  \nonumber \\
        & + & \{\xi^{g-1} + \xi^{g-2} + ........+ \xi^{2} + \xi + 1\} \\
        \end{eqnarray}
This can be recast as
\begin{eqnarray}
        \tau(g)& = & g + (g-1)\xi + (g-2)\xi^{2} + ...... 2\xi^{g-2} +
\xi^{g-1}
\end{eqnarray}

This series can be summed to yield eq. (12)

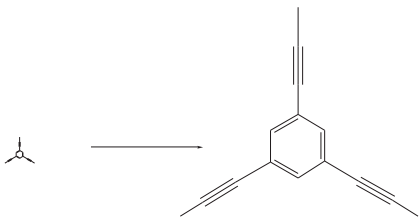
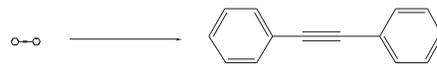
\begin{figure}
\caption{Extended Phenylacetylene Dendrimers with increasing total
number of generations.}
\label{Fig. 1}
\end{figure}

%\begin{figure}
%\caption{The MFPT for the linear potential with both types of
%random energy fluctuations (indistinguishable from each other)
%compared to that of the pure system (in the linear regime). }
%\label{Fig. 2}
%\end{figure}

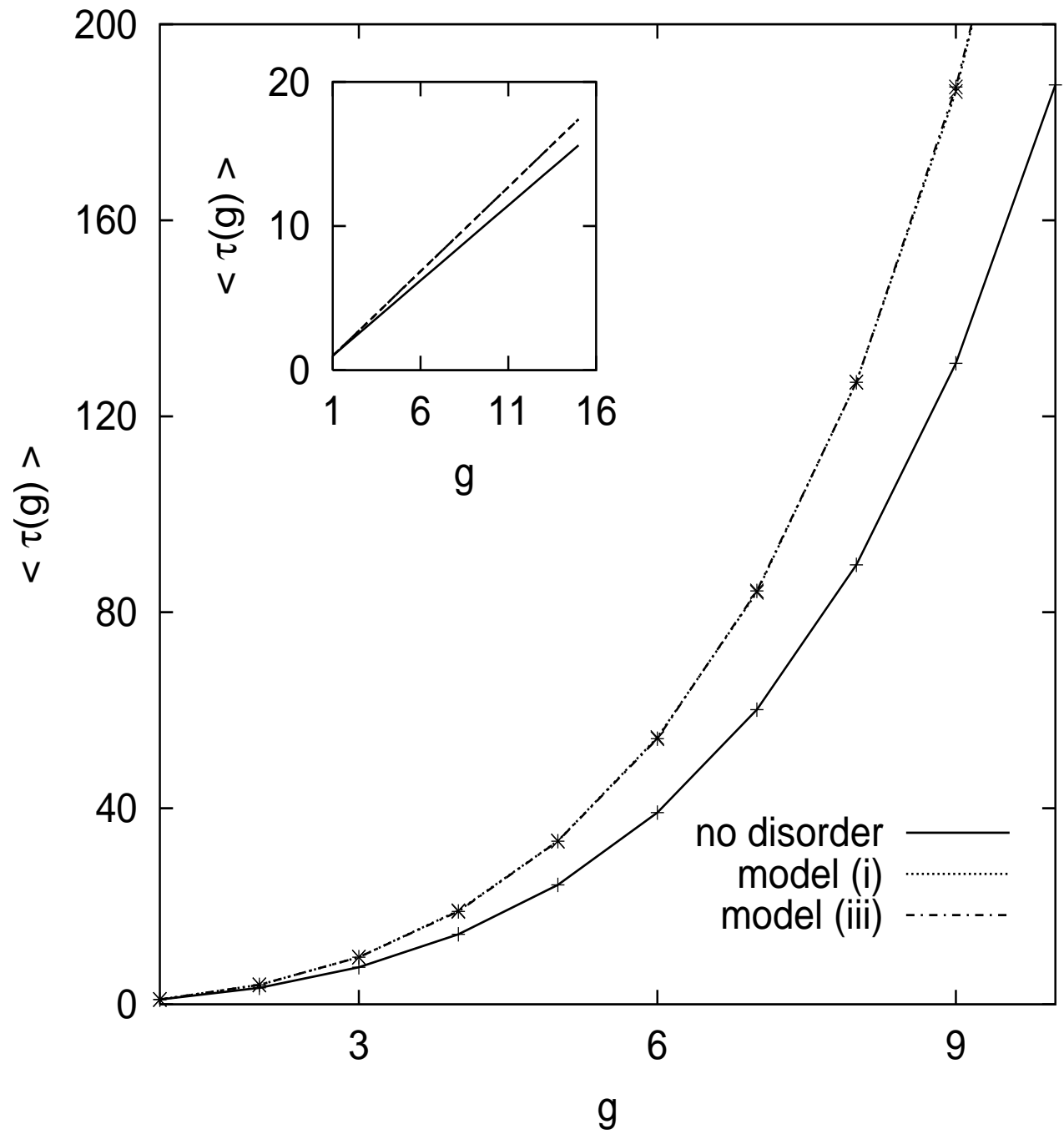
\begin{figure}
\caption{The MFPT for the linear potential with both types of
random energy fluctuations (indistinguishable from each other)
compared to that of the pure system (in the exponential regime).
Inset: the same in the linear regime }
\label{Fig. 2}
\end{figure}

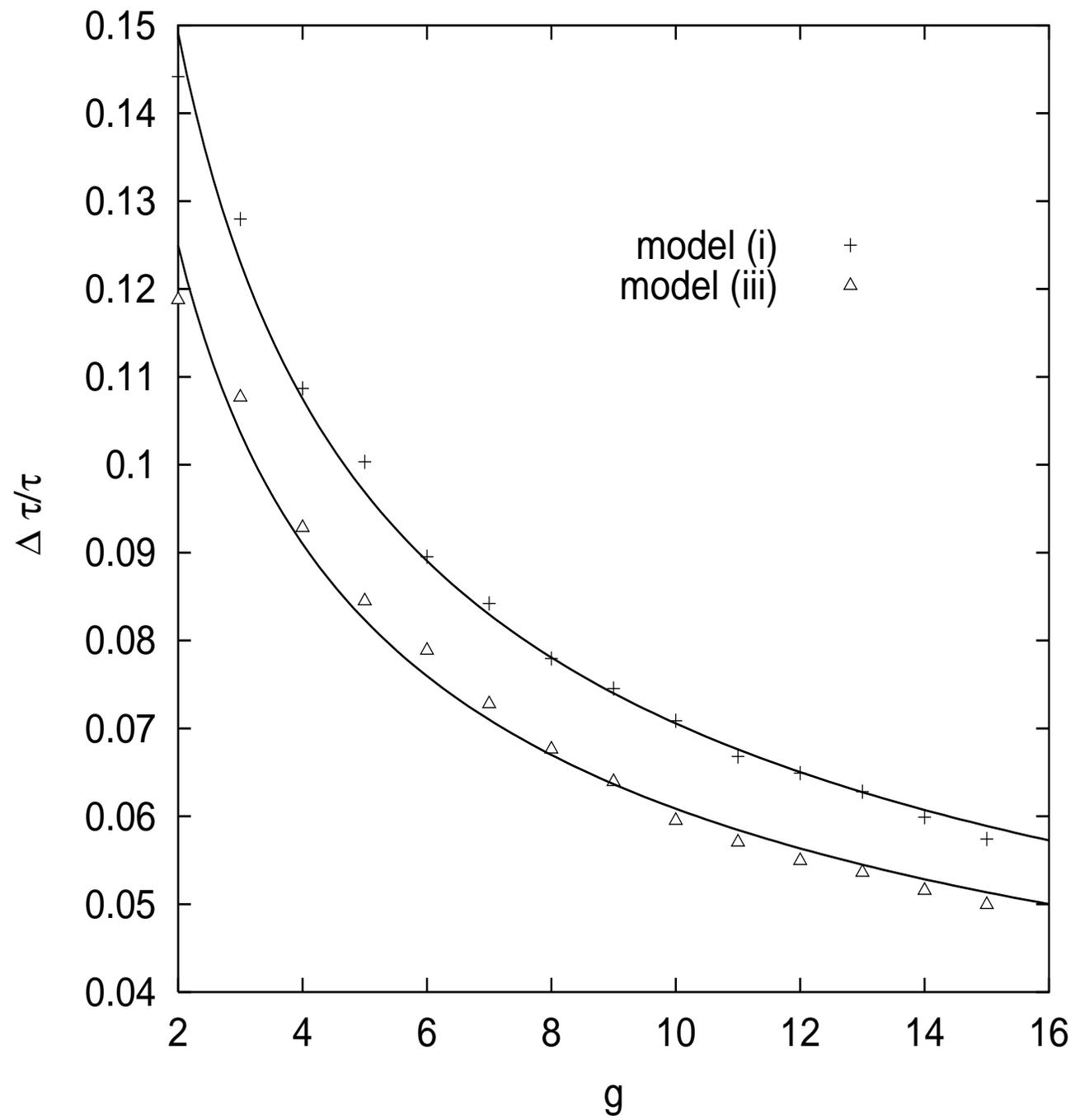
\begin{figure}
\caption{The relative fluctuation in the MFPT for both types of
random energy disorder (intergenerational and intersegment) with
a linear potential (in the linear regime). }
\label{Fig. 3}
\end{figure}

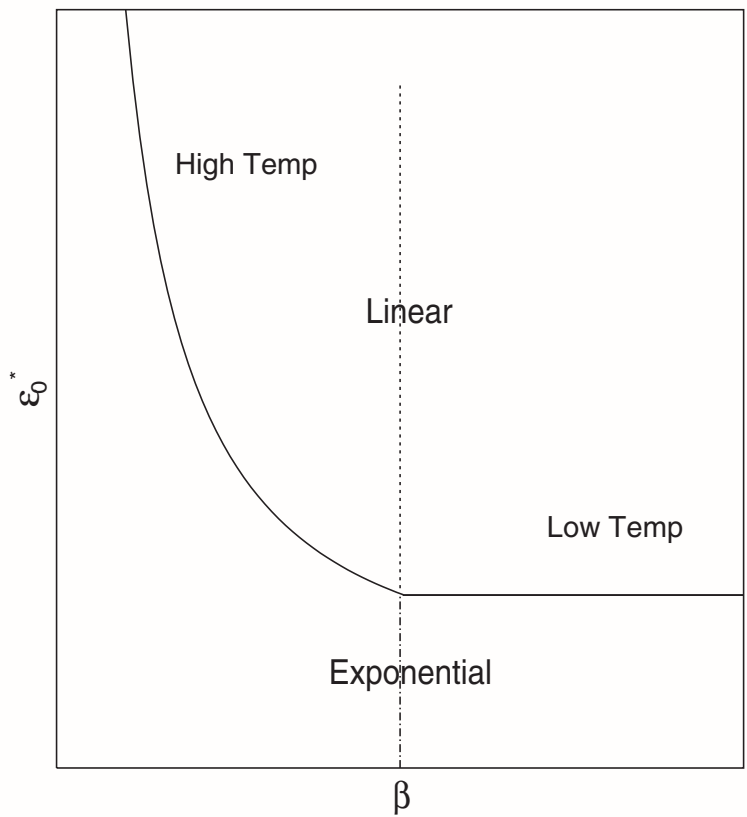
\begin{figure}
\caption{The ($\epsilon_{0},T$) phase diagram for the replica symmetry
breaking transition}
\label{Fig. 4}
\end{figure}

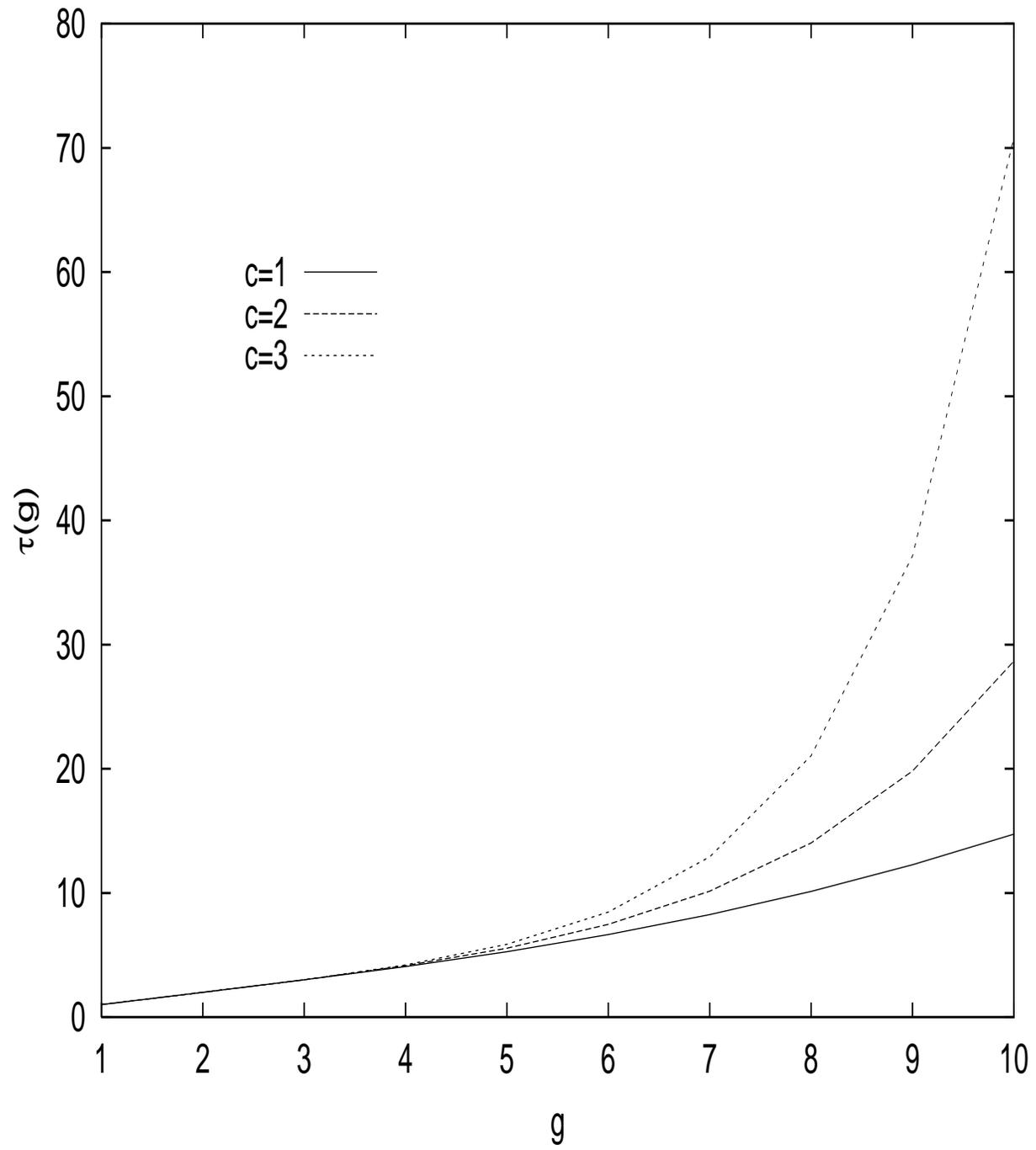
\begin{figure}
\caption{The MFPT {\it vs} $g$ for the ordered TDHF nonlinear potential for
different branching ratios of the tree}
\label{Fig. 5}
\end{figure}

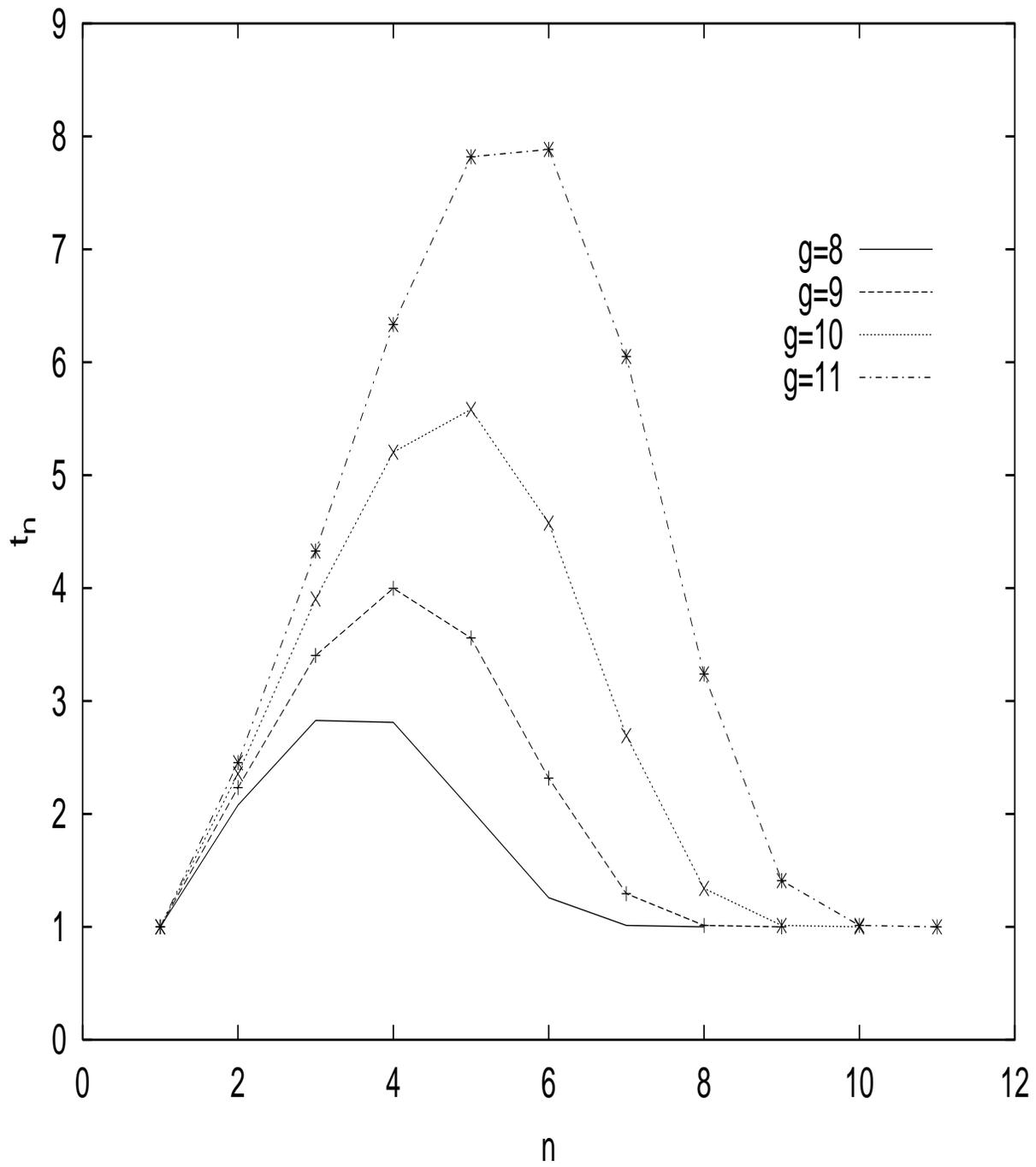
\begin{figure}
\caption{The MRT {\it vs} $n$ for the ordered TDHF potential, for different
values of $g$.}
\label{Fig. 6}
\end{figure}

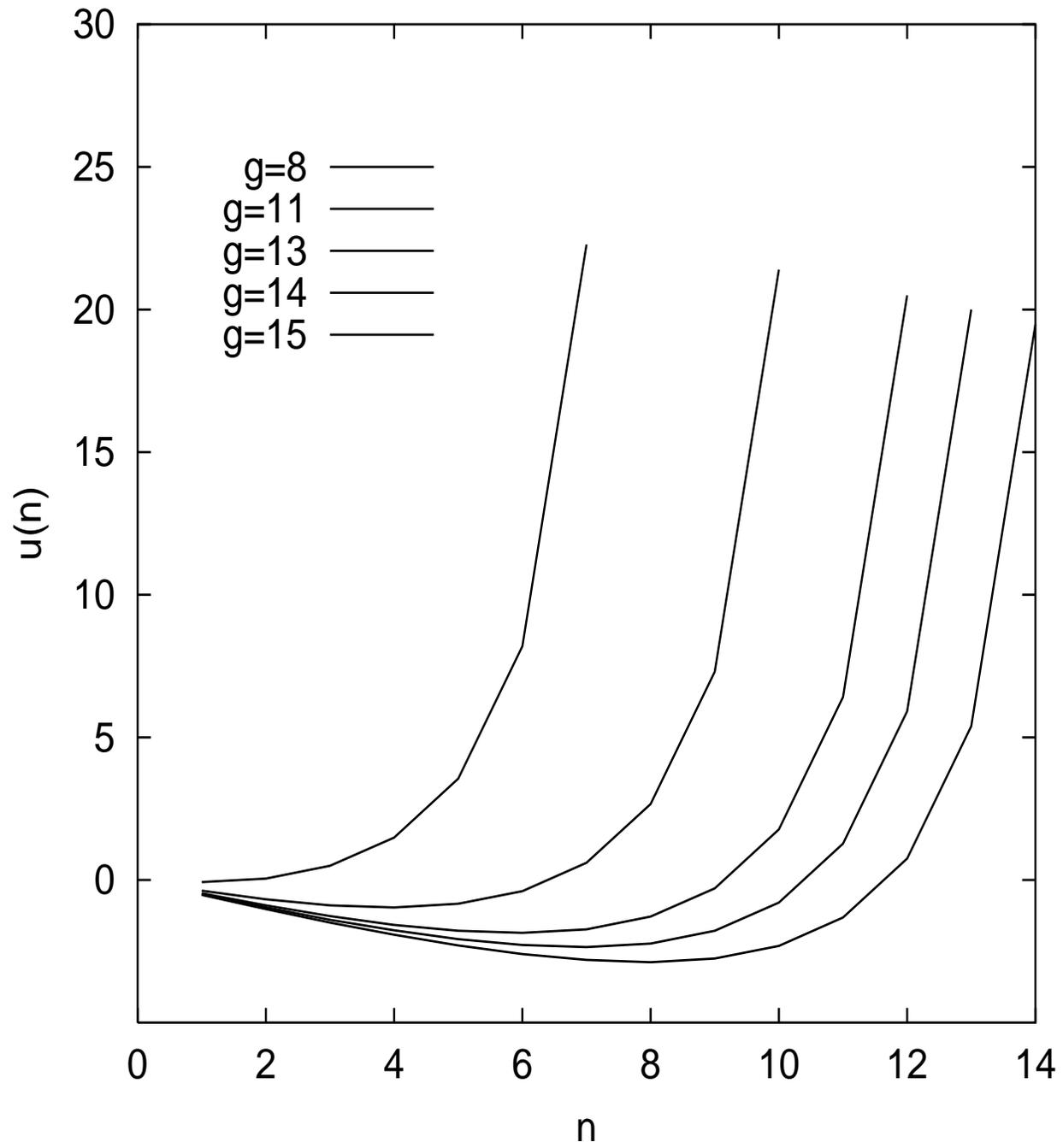
\begin{figure}
\caption{The free energy $u(n)$ for the ordered TDHF nonlinear
potential are plotted for different values of $g$.}
\label{Fig. 7}
\end{figure}

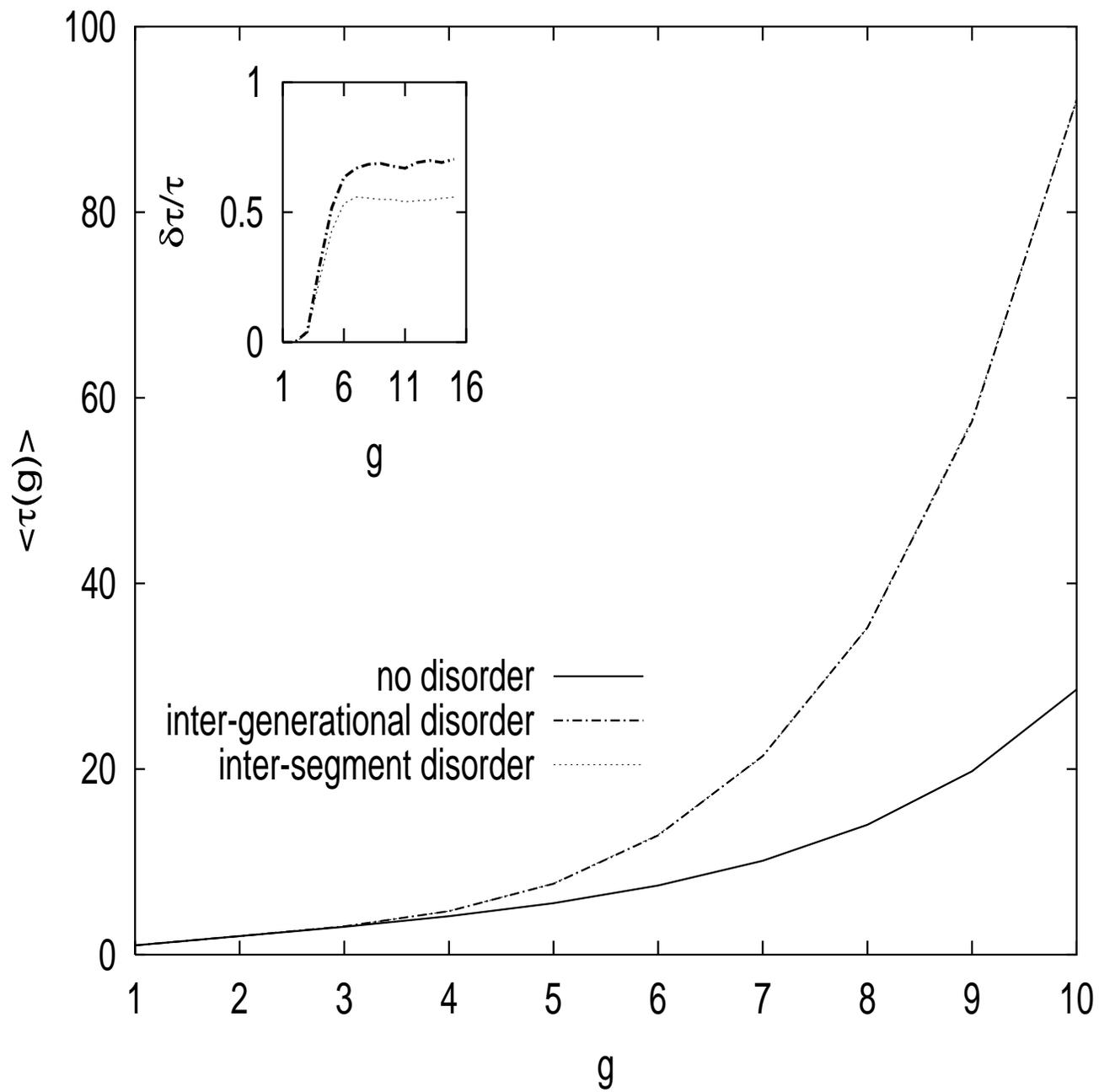
\begin{figure}
\caption{The MFPT for the TDHF potential with both types of
random energy (indiscernible from each other)
compared to that of the pure system. ({\it Inset:} Their relative $rms$
variations $vs$ $g$).}
\label{Fig. 8}
\end{figure}

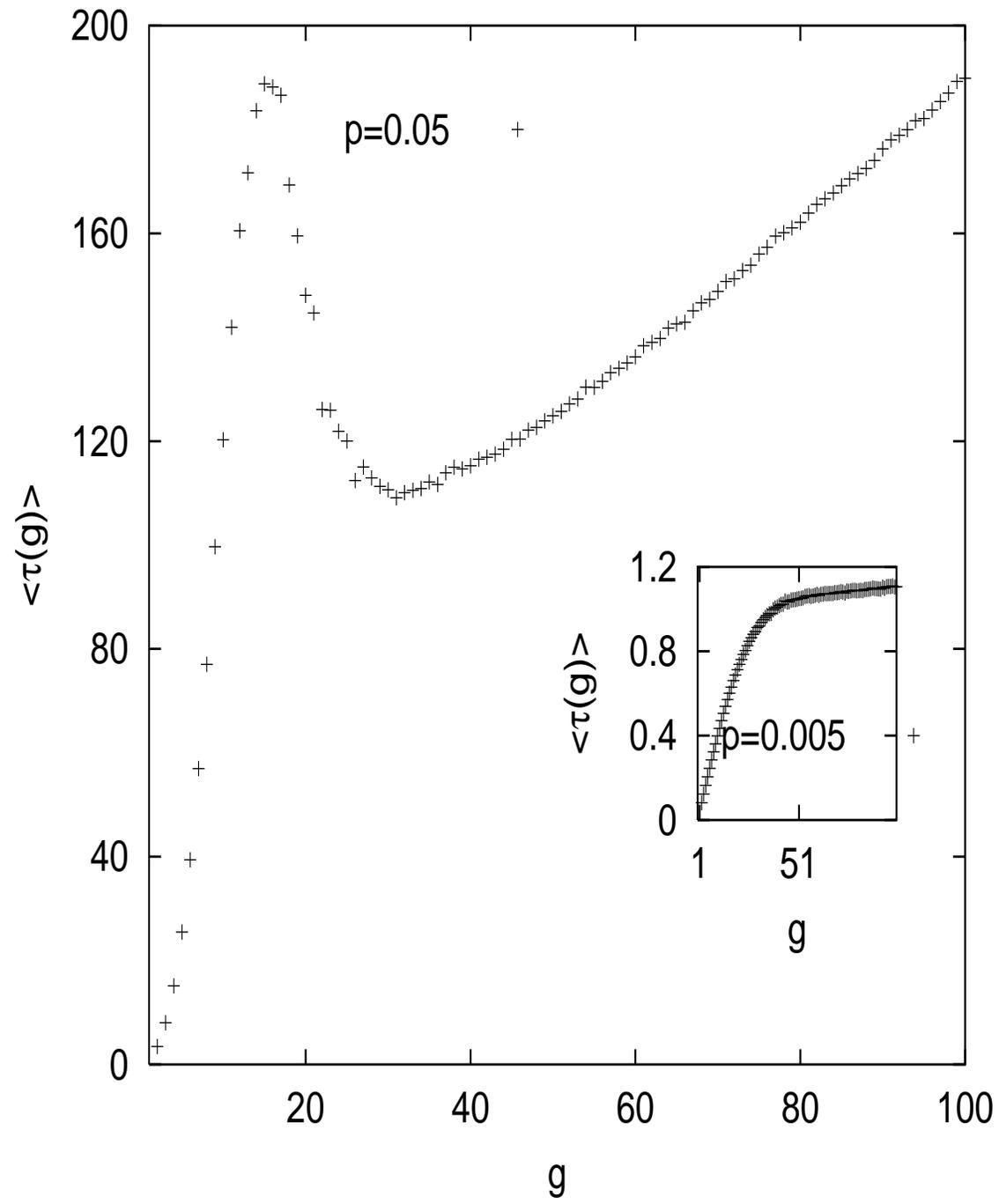
\begin{figure}
\caption{The MFPT {\it vs} $g$ for the small world model (exponential
regime). Inset: same for the linear regime.}
\label{Fig. 9}
\end{figure}

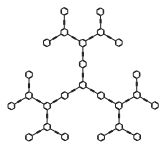
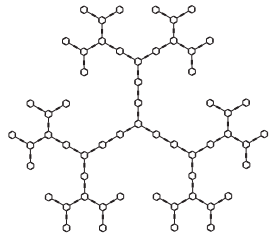
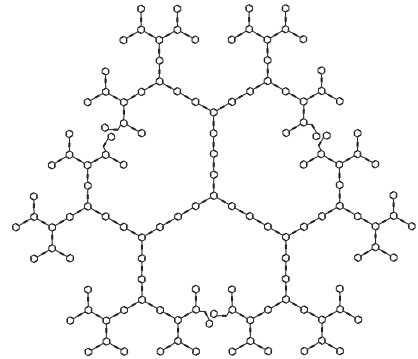
\end{document}